\documentclass[twocolumn]{IEEEtran}
\usepackage[cmex10]{amsmath}
\usepackage{enumitem}
\usepackage{graphicx}
\usepackage{color}
\usepackage{amsmath}
\usepackage{mathtools}
\usepackage{multicol}
\usepackage{multirow}
\usepackage[english]{babel}
\usepackage{blindtext}
\usepackage{algorithm}
\usepackage{algorithmic}
\usepackage{balance}
\usepackage{amsfonts}
\usepackage{bm}
\usepackage{stfloats}
\usepackage{subfig}
\usepackage{amsthm}
\usepackage{amssymb}
\usepackage{setspace}
\usepackage[nosort]{cite}
\usepackage{CJK}
\usepackage{cite}
\usepackage{caption}
\usepackage{comment}
\usepackage{array,multirow}
\usepackage{graphicx}

\theoremstyle{plain}

\usepackage[table]{xcolor}

\begin{document}

\title{UAV Swarm-Enabled Aerial Reconfigurable Intelligent Surface}
\author{\IEEEauthorblockN{Bodong Shang, Rubayet Shafin, and Lingjia Liu}
\thanks{B. Shang and L. Liu are with Wireless@VT, The Bradley Department of ECE, Virginia Tech, Blacksburg VA, 24061, USA. R. Shafin is with Standards and Mobility Innovation Lab at Samsung Research America, Plano, TX 75023, USA. The corresponding author is L. Liu (ljliu@ieee.org)}
\thanks{This work has been submitted to the IEEE for possible publication.
Copyright may be transferred without notice, after which this version may no longer be accessible.}
}
\maketitle

\begin{abstract}
Reconfigurable intelligent surface (RIS) offers tremendous spectrum and energy efficiency in wireless networks by adjusting the amplitudes and/or phases of passive reflecting elements to optimize signal reflection.
With the agility and mobility of unmanned aerial vehicles (UAVs), RIS can be mounted on UAVs to enable three-dimensional signal reflection.
Compared to the conventional terrestrial RIS (TRIS), the aerial RIS (ARIS) enjoys higher deployment flexibility, reliable air-to-ground links, and panoramic full-angle reflection.
However, due to UAV's limited payload and battery capacity, it is difficult for a UAV to carry a RIS with a large number of reflecting elements. 
Thus, the scalability of the aperture gain could not be guaranteed.
In practice, multiple UAVs can form a UAV swarm to enable the ARIS cooperatively.
In this article, we first present an overview of the UAV swarm-enabled ARIS (SARIS), including its motivations and competitive advantages compared to TRIS and ARIS, as well as its new transformative applications in wireless networks.
We then address the critical challenges of designing the SARIS by focusing on the beamforming design, SARIS channel estimation, and SARIS's deployment and movement.
Next, the potential performance enhancement of SARIS is showcased and discussed with preliminary numerical results.
Finally, open research opportunities are illustrated.
\end{abstract}

\section{Introduction}
The fifth-generation (5G) cellular network aims to achieve 10 Gbit/s peak data rate by having greater bandwidth, deploying denser networks, and multiplying the antenna links' capacity~\cite{7894280}.
However, the improved system performance comes at the cost of increased capital expenditures and operating expenses due to the enormous energy consumption generated by active hardware components.
In future wireless networks, i.e., beyond-5G or the sixth-generation (6G) network, more spectrum and energy-efficient yet cost-effective technologies need to be developed.

Reconfigurable intelligent surface (RIS) has been introduced as a new technology to improve wireless networks' spectrum and energy efficiency~\cite{cui2014coding,8796365}.
RIS is a planar surface which comprises large number of low-cost passive reflecting elements. By adjusting the amplitudes and phase shifts of the reflecting elements, RIS can achieve fine-grained reflection-beamforming.
In addition, with full-duplex mode of operation and with no noise-addition characteristics, RIS is more spectrum-efficient than the conventional relay technology, where the latter usually operates in half-duplex mode and has noise added to the signal at each relay node.
Moreover, RIS is envisioned to have no radio frequency (RF) chains, making it more energy efficient compared to massive multiple-input multiple-output (MIMO) system, which typically has tens or hundreds of RF chains. Therefore, RIS can be deployed in wireless networks to achieve significantly improved overall system performance.

Instead of being limited to only  terrestrial deployment, wireless networks are gradually evolving to air-ground integrated networks to achieve ubiquitous wireless connectivity and upgraded network capacity.
Recently, unmanned aerial vehicle (UAV) has attracted significant attention in wireless communication.
Due to the agility and mobility, UAVs can be quickly deployed in hotspots or disaster regions to support reliable communication, resorting to its line-of-sight (LoS)-dominated connections in the air-to-ground channels~\cite{8875722}.
As an aid to this, RIS can be mounted on UAVs to enable aerial RIS (ARIS) to achieve three-dimensional (3D) signal reflection.
Such ARIS is not restricted to the ${180^ \circ }$ half-space reflection, but instead, it provides a ${360^ \circ }$ panoramic full-angle reflection~\cite{lu2020enabling}.
With UAV's flight ability in 3D space, ARIS is more flexible in deployment than the conventional terrestrial RIS (TRIS), which is usually deployed on facades of a building or at a dedicated site.
Attaining an appropriate place for TRIS installation would not be easy in practice due to excessive site-rent and urban landscape impact.
Moreover, ARIS's cascaded reflection channel is more desirable than TRIS's, which shows the potential to improve system performance further.
From the above discussion, it is practically appealing to investigate the combination between UAVs and RISs.

Nevertheless, the number of reflecting elements on a single UAV is constrained due to UAV's limited payload, battery capacity, and flight flexibility.
Although RIS is generally lightweight and has conformal geometry, the large aperture gain of a single UAV-enabled ARIS would not be guaranteed.
Therefore, research on multiple cooperative UAVs (i.e., UAV swarm)-enabled ARIS is imperative, which can further improve the system performance.
In the following and in Fig. 1, the advantages of UAV swarm-enabled ARIS (SARIS) are summarized and compared with that of ARIS and TRIS.

\captionsetup{font={scriptsize}}
\begin{figure}[t]
\begin{center}
\setlength{\abovecaptionskip}{+0.2cm}
\setlength{\belowcaptionskip}{-0.0cm}
\centering
  \includegraphics[width=2.8in, height=2.8in]{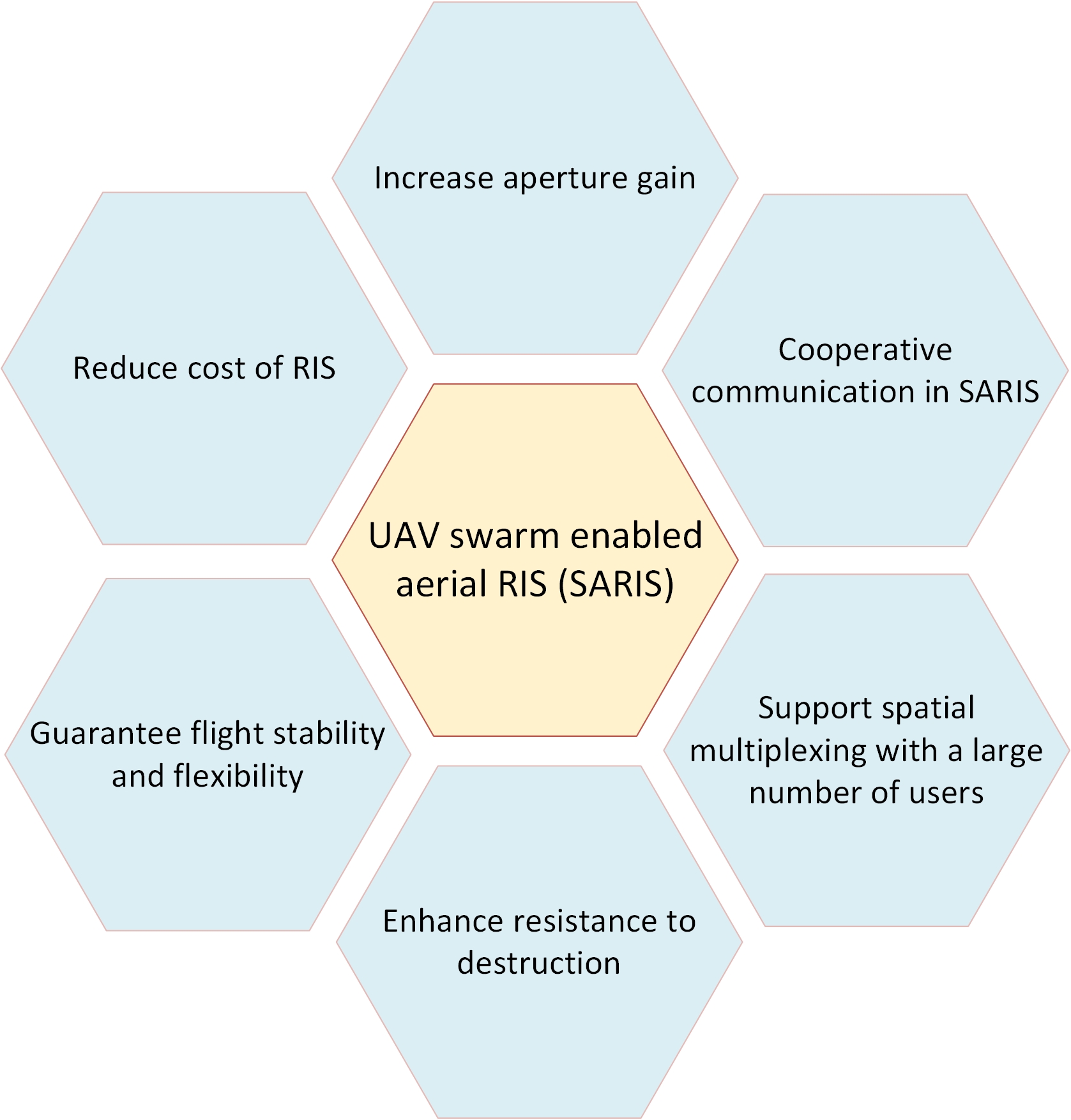}
\renewcommand\figurename{Figure}
\caption{\scriptsize Advantages of SARIS compared to ARIS and TRIS.}
\end{center}
\end{figure}

\captionsetup{font={scriptsize}}
\begin{figure*}[t]
\begin{center}
\setlength{\abovecaptionskip}{+0.2cm}
\setlength{\belowcaptionskip}{-0.0cm}
\centering
  \includegraphics[width=6.4in, height=4.2in]{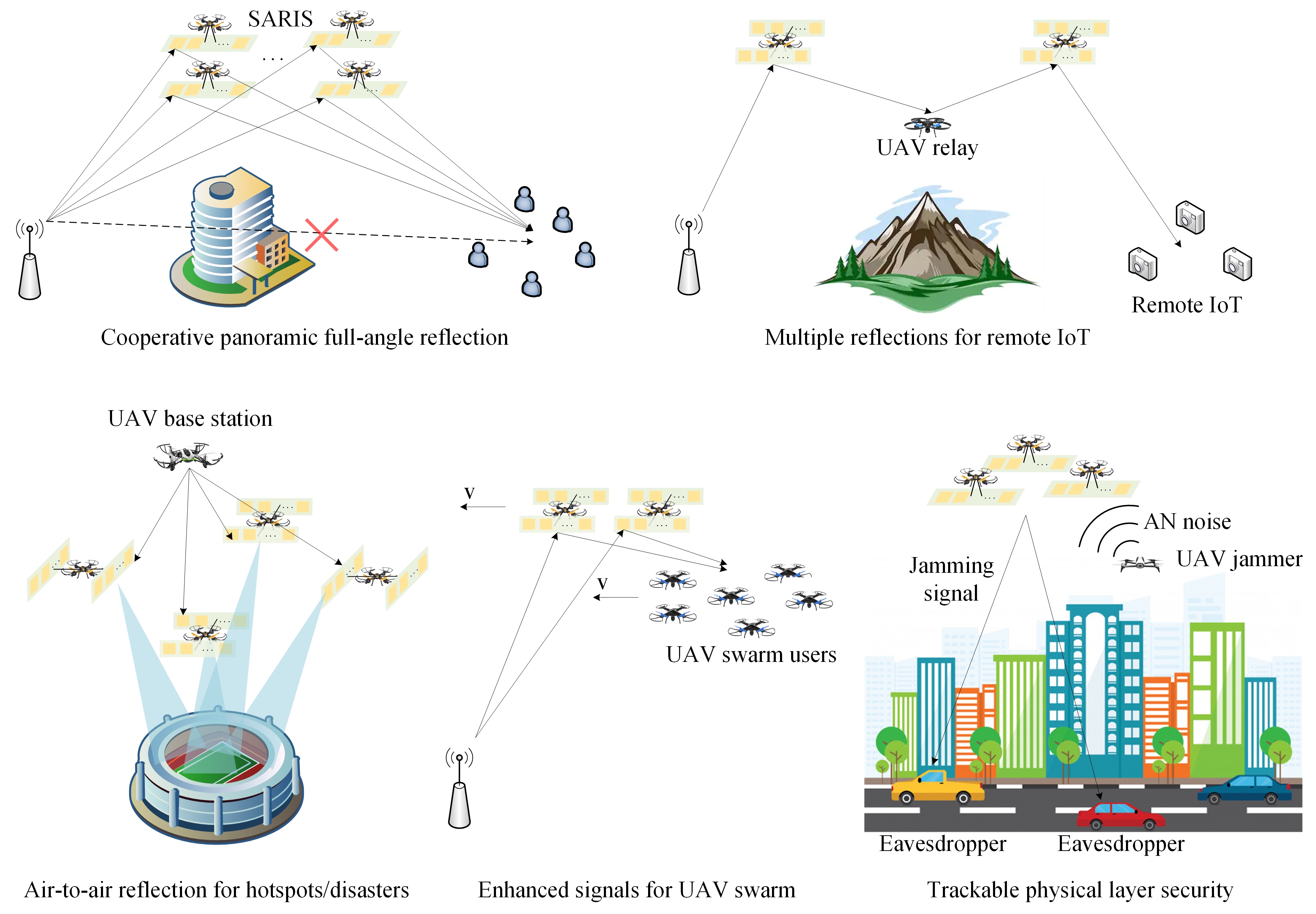}
\renewcommand\figurename{Figure}
\caption{\scriptsize Applications of UAV swarm-enabled aerial RIS in wireless network.}
\end{center}
\end{figure*}

\begin{itemize}
\item \textit{Increased aperture gain.}
In prior works~\cite{8811733}, it has been shown that RIS not only achieves a power gain of reflection-beamforming but also captures a power gain from collecting the received signal energy.
Under a single-user case with ARIS, the user's received power asymptotically increases in the order of $N^2$, where $N$ is the number of reflecting elements on ARIS.
If the number of UAVs, $L$, increases, user's received power approximately increases in the order of ${\left( {LN} \right)^2}$.
Accordingly, SARIS produces a notably enhanced aperture gain.
Moreover, compared to TRIS, SARIS can establish more reliable links to ground nodes with lower path loss exponent.

\item \textit{Cooperative communication.}
In future wireless networks, multiple UAVs can form a UAV swarm to complete a mission collaboratively in civil and military applications, such as surveillance, video streaming, and battlefield monitoring.
UAVs with different functionalities (e.g., ARIS, UAV base station (BS), UAV relay, and UAV user) can therefore cooperate with each other to provide reliable wireless communication.

\item \textit{Support for spatial multiplexing with a large number of users.}
In single UAV-enabled ARIS, the LoS links between the BS and ARIS result in low-rank MIMO channels, which usually cannot support multi-user spatial multiplexing.
In UAV swarm-enabled SARIS, multiple UAVs at  different positions, provide a rich scattering environment with diverse angles of arrival (AoA) at the BS.
Therefore, the formed MIMO channels have high rank to support spatial multiplexing with a large number of users.

\item \textit{Enhanced resistance to destruction.}
ARIS with a single UAV is vulnerable to the destruction resulted from attacks or malfunctions.
Consequently, if ARIS is disabled, reliable wireless connections will not be continuously guaranteed, and thus users will  suffer an outage.
On the contrary, SARIS holds an enhanced resistance to the destruction in operation based on multiple UAVs equipped with RIS.
Thus, a more robust aerial platform enabling RIS in the sky can be achieved by SARIS.

\item \textit{Guaranteed flight stability and flexibility.}
It is inconvenient for a UAV to guarantee flight stability and flexibility if it carries numerous reflecting elements, especially under bad weather conditions or air turbulence.
The UAV would consume more energy for an increased payload, which reduces its lifetime under the limited battery capacity.
However, when there are multiple UAVs, each UAV can carry fewer reflecting elements to ensure its flexible movement and prolonged lifetime.
Moreover, SARIS can track moving vehicles in vehicular networks to support reliable communication with its flexible movement.

\item \textit{Reduced cost of RIS.}
A large RIS consists of enormous integrated electric components resulting in an increased production cost.
Furthermore, the internal control of a large RIS is complicated.
Therefore, multiple moderate-sized RISs can reduce the production and control costs compared to a single large RIS.
\end{itemize}

\section{Applications of SARIS in Wireless Networks.}
In this section, new transformative applications of SARIS in wireless networks are illustrated, as depicted in Fig. 2.

\subsection{Cooperative Panoramic Full-Angle Reflection}
When users are in a dead zone where an obstacle severely blocks the direct link to the BS, multiple UAVs can cooperatively serve as a SARIS to provide ${360^ \circ }$ panoramic full-angle reflection for ground users.
Note that TRIS on the facade of a building can support at most ${180^ \circ }$ half-space reflection, i.e., both the source and destination nodes are on the same side of TRIS.
In SARIS, each reflecting element can independently control the amplitude and/or phase of the incident signal.
This constructively adds the desired signals and suppresses the interfering signals at ground users, and it can provide a significant aperture gain for multiple users.
In this application, SARIS's deployment is essential to the system performance, where the distribution of ground nodes and blockages needs to be considered.

\subsection{Multiple Reflections for Remote Internet-of-Things (IoT)}
 Remote areas, such as forest, vast ocean, volcanic, and other arduous environments, are difficult to be covered by the current cellular network.
The IoT devices may be widely deployed for specific tasks that require data transmissions, e.g., the fusion of sensing data of high-definition sound and video information.
SARIS can be deployed in future wireless networks to provide ubiquitous wireless connectivity with reliable data transmissions for remote IoT.
According to the product-distance-based path loss (i.e., doubled path loss) model in RIS communication, purely multiple SARIS reflections are inefficient due to the dramatically increased signal attenuation.
Fortunately, UAVs with different functionalities (e.g., ARIS, UAV relay) can cooperatively support data transmissions for remote IoT as a UAV swarm.
For example, an ARIS can reflect the signal to a UAV relay, and then the UAV relay decodes and forwards the signal to another ARIS shown in Fig. 2.
In this way, a reliable wireless link is established in demand between BS and remote IoT devices.

\subsection{Air-to-Air Reflection for Hotspots/Disaster Areas}
In cellular networks, ground small cell BSs (SBSs) are usually deployed in areas of high user density (termed hotspots), such as downtown areas, stadiums, and squares.
However, the distributions of hotspot users are usually dynamic in practice.
In this scenario, SARIS can provide wireless communications for hotspot users, shown in Fig. 2.
For example, wireless signals are transmitted from UAV BS and reflected by SARIS to hotspot users with significant aperture gain.
Due to the reliable air-to-ground channels, the signal strength under such air-to-air reflection is improved significantly compared to conventional SBSs.
Moreover, UAVs' positions can be adjusted according to ground users' real-time distribution to further improve system performance.
In addition, such an air-to-air reflection system can be deployed in regions of disasters, where the terrestrial infrastructure is disabled.

\subsection{Signal Enhancement in UAV Swarm}
In some cases, UAV swarm needs to connect to BS for information exchange, such as, the flight control messages, the machine learning model for autonomous flight, and the offloaded computational tasks.
Therefore, the reliable wireless communication between the UAV swarm and BS should be ensured.
With this aim, the UAV-enabled ARIS, or SARIS can fly synchronously alongside the UAV swarm users to provide the reliable reflected signal for UAV swarm users, as shown in Fig. 2.
In this application, UAVs' velocity and movement need to be considered in the beamforming design.

\subsection{Tractable Physical Layer Security}
Vehicle-to-vehicle (V2V) communication allows vehicles' connections in proximity.
Secured vehicular communication becomes critical in data-oriented applications for certain services such as control messages, maneuver command, and file transfers.
However, as vehicles are mobile, it is challenging to secure V2V communication from a physical layer perspective with terrestrial infrastructure.
Specifically, the unknown and varied small-scale fading components in vehicular networks impede the channel acquisition for TRIS.
With SARIS's help, as shown in Fig. 2, a UAV jammer sends an artificial noise (AN)-based jamming signal to SARIS, and then SARIS reflects the jamming signal to the potential eavesdroppers.
Moreover, due to the guaranteed flight stability and flexibility, SARIS can track the potential eavesdropper or follow the legitimate vehicle to provide continuously secured communication.

\section{Challenges of Designing SARIS}
In this section, we discuss the main challenges behind practical realization of SARIS, including beamforming, channel estimation, deployment, and movement of SARIS.

\subsection{Beamforming Design}
With massive passive reflecting elements, SARIS serves multiple users with significant aperture gain.
Note that the scheduling of users in a time-slot affects the system performance and the subsequent beamforming design.
Users' scheduling is related to users' priority, the objective of formulated problem, and user's mutual interference, etc.
Moreover, if SARIS moves dynamically, users' scheduling over different time-slots needs to be designed based on SARIS' trajectory.
Given users' scheduling in a time-slot, the active and passive beamforming are designed accordingly.

\subsubsection{System performance versus algorithmic complexity}
For a general multi-user setup, the active and passive beamforming optimization problem often arises as a multiple-ratio fractional programming (FP) problem.
This is because the system design usually involves multiple signal-to-interference-plus-noise ratio (SINR) terms.
Multiple-ratio FP problems are non-convex problems in general.
In \cite{8811733}, the authors used the semidefinite relaxation (SDR) technique to solve the non-convex passive beamforming problem.
However, SDR may not be suitable for large-scale SARIS with massive reflecting elements since the number of involved variables is quadratic in the number of reflecting elements.
In \cite{8314727}, a quadratic transform was used to facilitate a non-convex passive beamforming problem as a sequence of convex problems.
In each iteration, the active and passive beamforming coefficients can be optimized alternately.

For achieving low algorithmic complexity, a time-slot can be divided into multiple sub-time-slots, and users are served in turn over sub-time-slots.
Under such a single-user setup, the active beamforming is obtained in closed form, i.e., maximum-ratio transmission.
Moreover, the passive beamforming is also obtained in closed form to align and combine the user's signals, which reduces the algorithmic complexity.
Therefore, there is a trade-off between system performance and algorithmic complexity.
In addition, stochastic geometry can be used to analyze the optimal deployment of SARIS even under the randomly distributed users, which is accomplished by numerical calculations instead of extensive and time-consuming Monte Carlo simulations.

\subsubsection{Artificial intelligence (AI) for SARIS beamforming design}
To further reduce the algorithmic complexity, deep learning techniques can be used to obtain the active and passive beamforming matrix based on channel state information (CSI).
For example, a deep neural network (DNN) can be trained to map CSI to the active/passive beamforming according to a complete offline sample library established by practical operation or simulations.

In some UAV systems, it is difficult to obtain the channel model or users' channel information due to the unknown and intricate environment, e.g., various buildings and obstacles.
Deep reinforcement learning (DRL) can be utilized to obtain efficient beamforming design by observing rewards from the environment \cite{9110869}.
However, the DRL-based approach is not suitable for highly dynamic networks where users and obstacles are moving since DRL needs time to converge to a stable solution given the network topology.

{\rowcolors{1}{blue!60!green!5}{blue!60!green!15}
\begin{table*}[t]
\captionsetup{font={normalsize}}
\caption{Design challenges and potential solutions for SARIS.}
\centering
\begin{tabular}{ | m{12em} | m{6cm}| m{6cm} | }
\hline
\textbf{SARIS Implementation} & \textbf{Challenges} & \textbf{Potential solutions}  \\
\hline
Beamforming design & a) Increased overhead for UAV swarm; b) The impact of UAV wobbling on accurate CSI acquisition; c) Computation complexity of iterative algorithm. & a) Cooperative information transfer in UAV swarm; b) Robust beamforming design with CSI error models; c) Single user scheduling, or deep learning methods for multi-user passive beamforming. \\ \hline
Channel estimation & a) High channel estimation overhead; b) Practical limitations on the number of RF chains; c) Pilot contamination due to strong LoS paths between the UAV and multiple BSs. & a) Novel channel estimation protocol design tailored for SARIS; b) Overhead reduction by grouping adjacent reflecting elements forming sub-surfaces; c) Parametric channel estimation for mmWave systems. \\ \hline
Deployment and movement & a) Random distributions of UAVs and ground users; b) Deployment under uncertain LoS and NLoS connections in 3D space; c) Movement design under complicated mathematical model. & a) Spatial point process modeling and analysis; b) Stochastic geometry modeling and analysis to capture the large-scale fading component; c) Reinforcement learning approach for obtaining a sub-optimal 3D trajectory. \\ \hline
\end{tabular}
\end{table*}}

\subsection{Channel Estimation for SARIS}

Channel estimation is one of the most critical steps for any RIS-assisted communication system.
The overhead is relatively high since channels corresponding to multiple links need to be estimated, i.e., the link between RIS and user, the link between RIS and BS, and the direct link between BS and the user. 
For a single user SARIS system where each user has one antenna, the number of channel coefficients that need to be estimated in a SARIS system is $(MNL+M)$, where $M$ is the number of antennas at the BS. 
For a multi-user SARIS system, this number goes to $KMNL+KM$, where $K$ is the number of users.

Another critical issue with massive MIMO channel estimation is the so-called pilot contamination (PC) that originated from the same pilots' re-use at adjacent cells.
In terrestrial massive MIMO networks, PC can be tackled using sufficiently large pilot re-use factors to ensure that neighboring BSs do not use the same pilot sequence. 
However, this method is not adequate for UAV communication since there are still strong LoS paths between the UAV and BSs, even if the BSs are quite far apart. 
The PC issue can also be addressed by employing coordination among the BSs.
However, due to the strong LoS paths between the UAV and the BSs, the number of BSs that need to cooperate for the UAV system is significantly larger than conventional BS cooperation, which incurs considerable overhead.
Therefore, efficient and low-cost channel estimation strategies need to be devised for the SARIS system.

\subsubsection{SARIS Channel Estimation Protocol} 
Assume that the $N$ reflecting elements in the RIS can be grouped into $N'$ sub-surfaces where each sub-surface contains $\Bar{N}$ elements. 
As such, $N'+1$ pilot tones are reserved to estimate the channel.
In general, channel estimation for the aggregated links (for the direct and reflected path) should be performed. 
Since the reflecting elements are passive, channel estimation needs to be done for the entire set of pre-designed reflection states at the RIS. Therefore, with $N'+1$ pilot symbols, the BS can estimate the direct and reflected channels corresponding to $N'+1$ reflection states. 
Next, the optimal reflection state is calculated based on the channel information for the data transmission. 
Finally, the optimal phase shift values are fed back to the RIS controller, which sets the recommended phase shifter setting for the highest data rate. 
Estimation accuracy improves with the number of pilot tones, and so does the estimation overhead. 
Hence, there is an inherent trade-off between finding the optimal phase shifter and the associated estimation overhead. 
This approach is more suitable for sub-6 GHz channels. 
Millimeter-wave (mmWave) channels can reduce the estimation overhead as described next.

\subsubsection{SARIS Channel Estimation for mmWave Systems} 
MmWave channels are much sparser than sub-6 GHz channels, i.e., the number of channel clusters and resolvable paths in mmWave channels are significantly fewer than those of sub-6 GHz channels. 
Therefore, using parametric channel modeling, the mmWave channels can be characterized by only a few channel parameters corresponding to each resolvable path. 
Such parameters include AoA, angles of departure (AoD), and channel gains. 
In contrast to the channel transfer function estimation based channel acquisition method for SARIS as explained in the previous sub-section, estimation overhead for parametric approach doesn't grow with the number of antennas either at the BS or the number of reflecting elements at RIS. 
Channels can be estimated by evaluating the path parameters followed by channel reconstruction based on parameters being assessed. 
In this regard, mmWave channel estimation for SARIS can be treated as a sparse signal recovery problem, and existing compressed sensing methods such as orthogonal matching pursuit can be deployed for channel estimation. 
Alternatively, BS can estimate the AoAs, and, based on the estimated angles, complex path gains can be calculated using the maximum likelihood method~\cite{shafin2018joint}.

\subsection{Deployment and Movement of SARIS}

\subsubsection{Deployment of SARIS}
The effective deployment of SARIS plays a significant role in system design.
Recall that RIS asymptotically achieves "square law" on the user's received power in the order of $N^2$ as the number of reflecting elements, $N$ ,goes to infinity \cite{8910627}.
However, the doubled large-scale path loss may severely deteriorate the system performance, which needs to be considered in the system design.
In the TRIS system, it is desirable to deploy RIS near either BS or user to reduce the doubled path loss, but this may not apply to SARIS.

The 3D position of SARIS not only determines the large-scale path loss but also changes the LoS and non-line-of-sight (NLoS) probabilities in air-to-ground channels.
Specifically, the LoS probability of air-to-ground connection depends on UAV's elevation angle and the environment parameters in specific scenarios, e.g., suburban, urban, dense urban, highrise urban, etc~\cite{6863654}.
In general, the LoS probability decreases with UAV's elevation angle.
Nevertheless, raising UAV's elevation angle increases the large-scale path loss due to a long communication distance.
Therefore, there is a trade-off between the doubled path loss and the excessive path loss of NLoS connections.
Moreover, UAVs and ground nodes' distributions also impact system performance.
SARIS's optimal 3D position can be investigated using stochastic geometry, which incorporates the large-scale path loss, LoS and NLoS probabilities, and UAVs' and ground users' distributions.
In addition, stochastic geometry can also be used to analyze the randomness and statistics of interference generated by other deployed TRIS.

\subsubsection{Movement of SARIS}
If SARIS serves multiple users in a period, optimizing SARIS's trajectory can improve system performance.
This is because UAVs can fly towards the scheduled users over time-slots to enhance the signal strength and increase the LoS probability.
For a UAV swarm, its formation can remain stationary, and one could optimize the trajectory of a reference point of the UAV swarm.
Besides UAVs' unified flight, a UAV swarm can be divided into multiple sub-swarms, and the trajectories of sub-swarms can be optimized to cooperatively serve multiple users while mitigating interference by adjusting their trajectories and beamforming strategies in the period.
It is worth noting that the signal reflection is sensitive to UAV wobbling which results from its movement and various environmental issues, such as random wind gusts.
As such, system performance may degrade due to signal misalignment.
Therefore, robust algorithms for beamforming with Quality-of-Service (QoS) guarantees are required.

In some cases, it is difficult to mathematically derive the optimal 3D trajectory of a UAV swarm due to the complicated problem formulation, which involves the LoS and NLoS-based air-to-ground channel model.
Since UAV's 3D trajectory design is relevant to a time sequence problem, reinforcement learning (RL) can be leveraged to learn a sub-optimal solution. 
In RL, the UAV swarm acts as an agent, learning its 3D trajectory from trial and mistake offline.
Multi-agent multi-task RL can be used to obtain solutions for multiple UAVs.
Meanwhile, each reflecting element's discrete phase shift facilitates a UAV to choose an action from a finite action space for passive beamforming, which can be included in the RL algorithm together with UAV's trajectory design.

\captionsetup{font={scriptsize}}
\begin{figure*}[t]
\begin{center}
\setlength{\abovecaptionskip}{+0.2cm}
\setlength{\belowcaptionskip}{-0.0cm}
\centering
  \includegraphics[width=4.5in,height=2.4in]{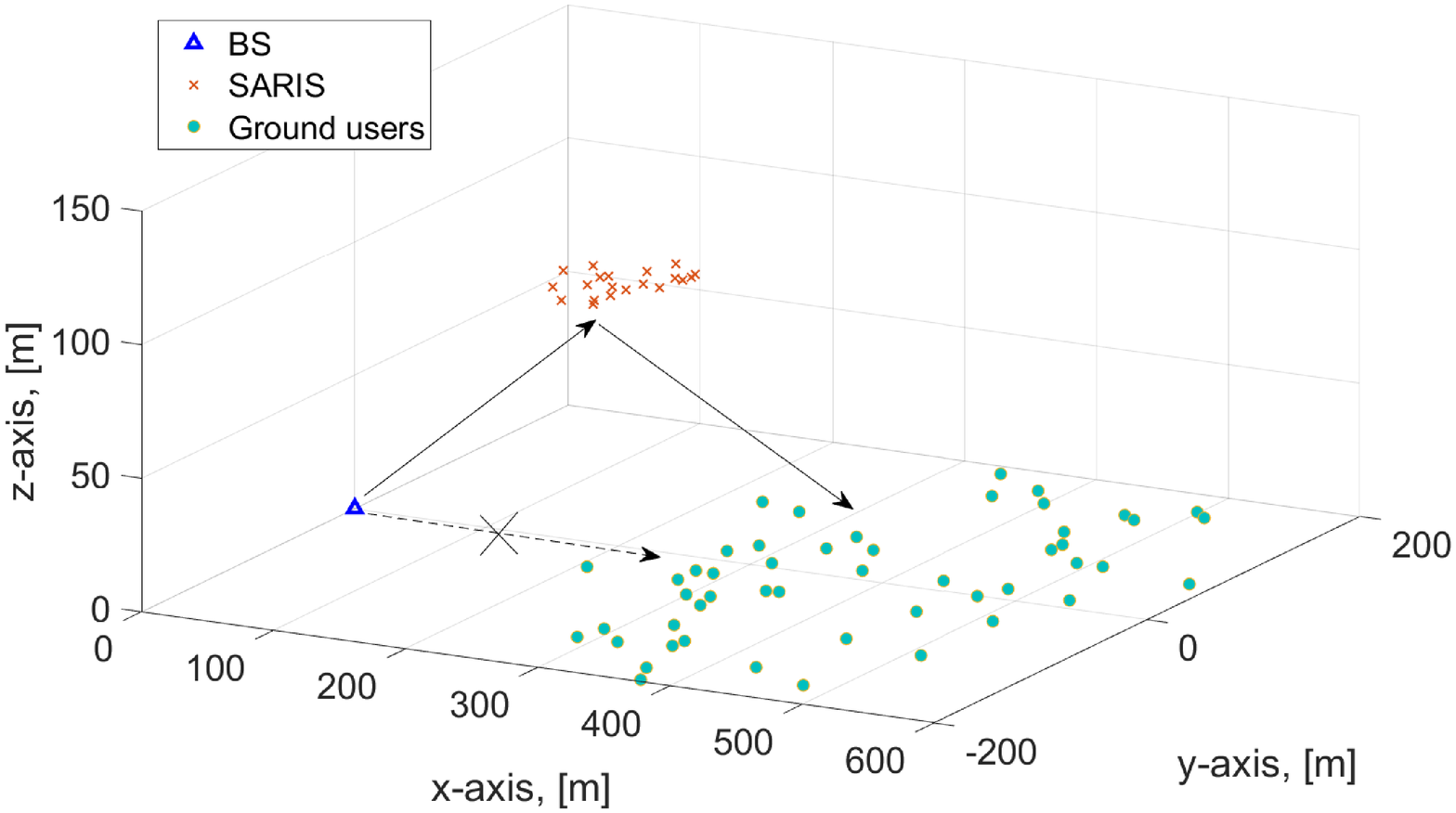}
\renewcommand\figurename{Figure}
\caption{\scriptsize Simulation setup.}
\end{center}
\end{figure*}

\captionsetup{font={scriptsize}}
\begin{figure}[t]
\begin{center}
\setlength{\abovecaptionskip}{+0.2cm}
\setlength{\belowcaptionskip}{-0.0cm}
\centering
  \includegraphics[width=3.4in, height=3.0in]{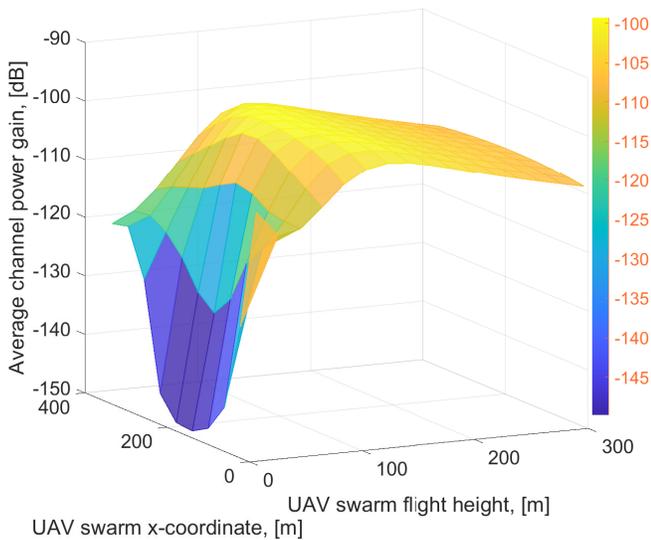}
\renewcommand\figurename{Figure}
\caption{\scriptsize Average channel power gain versus UAV swarm 3D position.}
\end{center}
\end{figure}

\begin{figure*}[t]
\setlength{\abovecaptionskip}{+0.2cm}
\setlength{\belowcaptionskip}{-0.2cm}
\centering
  \subfloat[Achievable rate versus $L$, where $R_A = 10 \text{ m}$, $R_U = 100 \text{ m}$.]
  {\includegraphics[width=3.2in,height=2.8in]{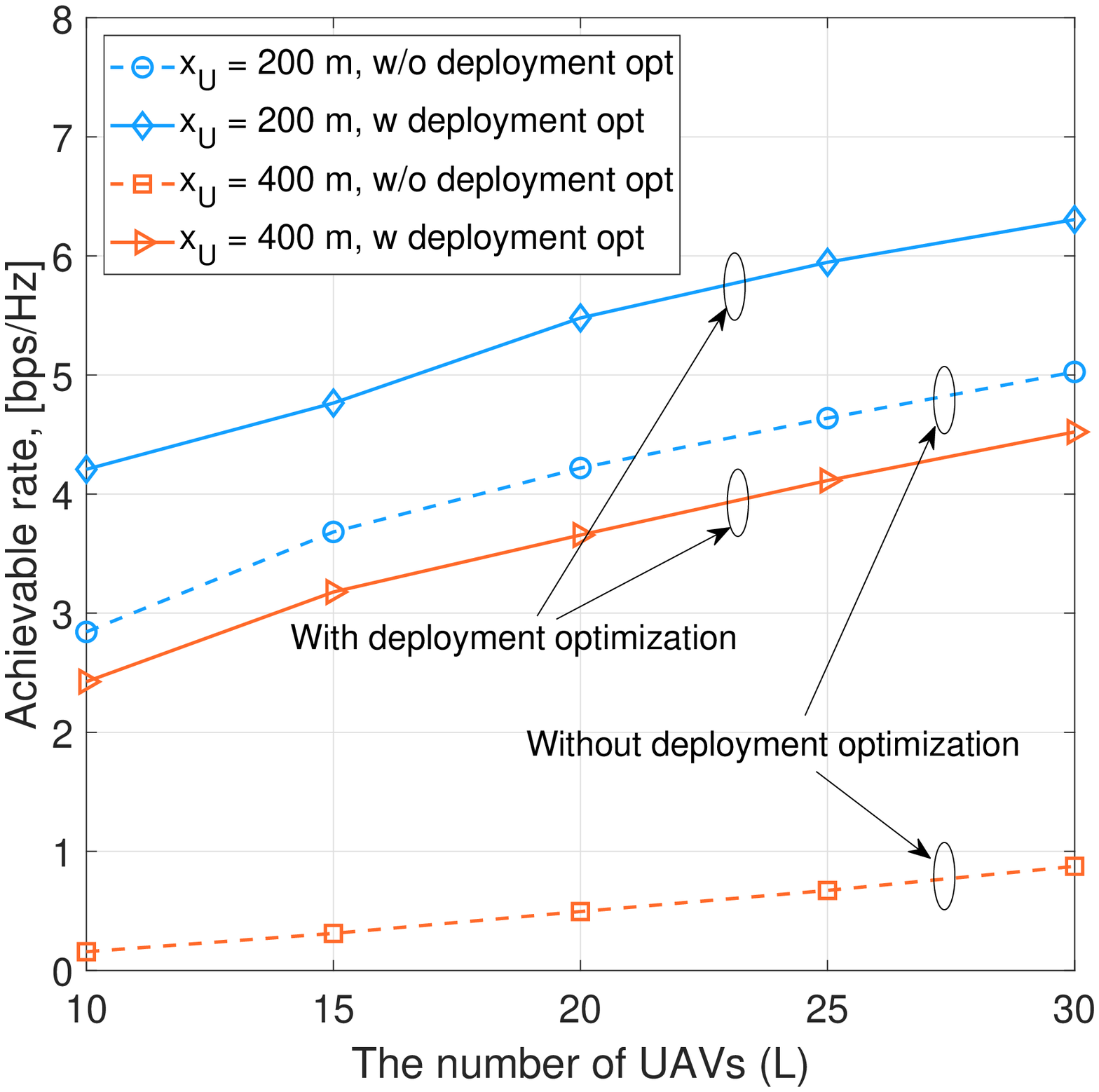}}
  \subfloat[Achievable rate versus $R_A$, where $x_U = 200 \text{ m}$, $L = 10$.]
  {\includegraphics[width=3.2in, height=2.8in]{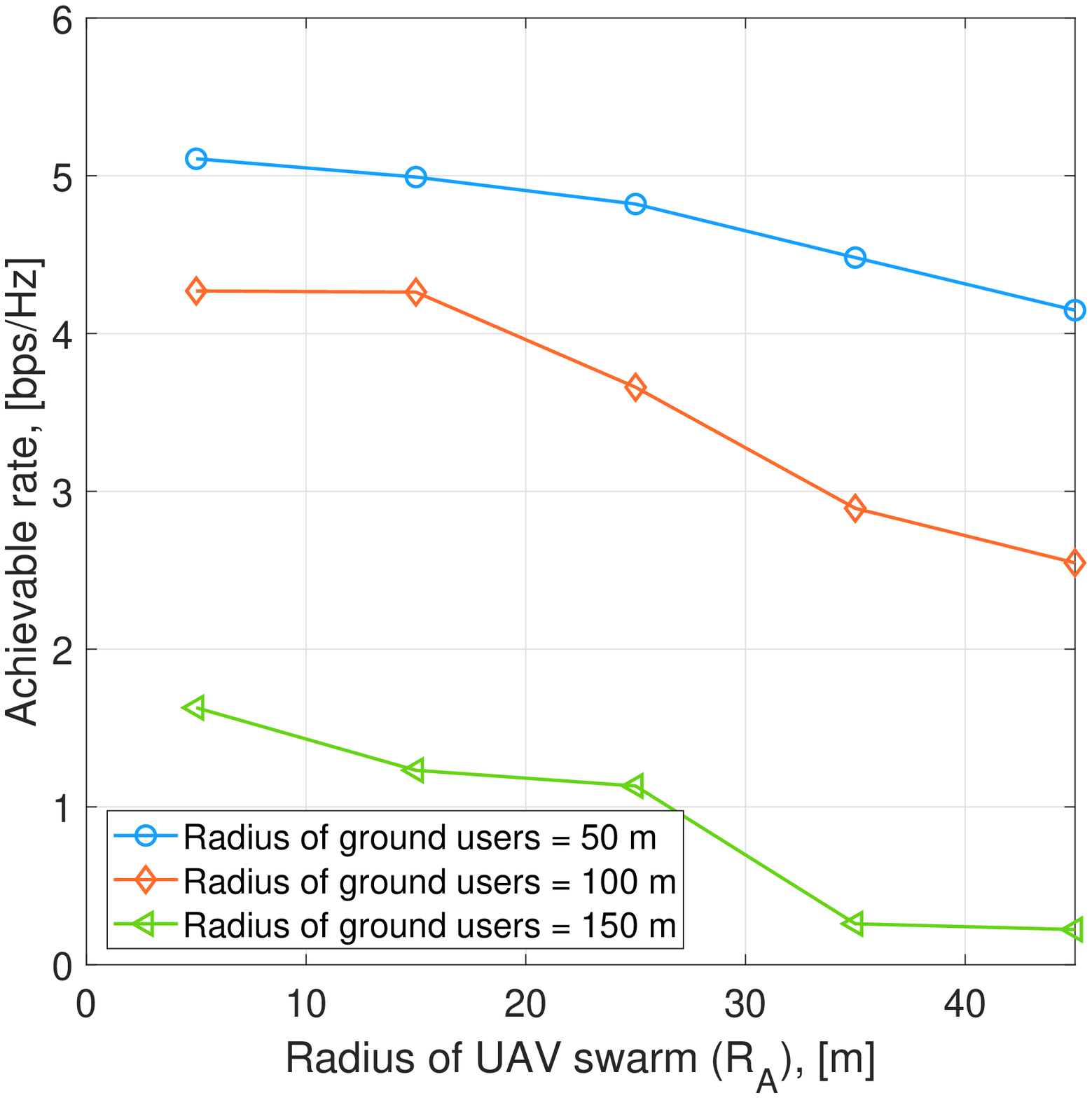}}
  \renewcommand\figurename{Figure}
  \caption{Achievable rate versus the number of UAVs and the radius of UAV swarm.}
\end{figure*}

\section{Simulation Results}
The simulation setup is shown in Fig. 3.
We assume that the BS serves the ground users with SARIS's help.
We consider 16 antennas at the BS and 20 reflecting elements on each UAV.
There are $L$ UAVs that fly at $H$ meters (m) from ground and are distributed based on a Matérn cluster process with a radius of $R_A$~\cite{haenggi2012stochastic}.
Ground users are distributed based on another Matérn cluster process with a radius of $R_U$.
Denote the distance between the BS and the center of users by $x_U$ m.
The air-to-ground path loss channel model is based on the modeling in \cite{6863654} under a dense urban environment.
The small-scale fading for NLoS components are independently drawn from the circularly symmetric complex Gaussian distribution with zero mean and unit variance.
The reflection efficiency of RIS is 0.9.
Simulation results are averaged over 1,000 experiments.
Finally, the noise power is -80 dBm~\cite{8910627}.

\subsection{Optimal Deployment of SARIS}
We consider that the SARIS is deployed to serve the ground users in Fig. 3, where $x_U = 200 \text{ m}$, $L = 10$, $R_A = 10 \text{ m}$, $R_U = 100 \text{ m}$.
We optimize the 3D position of the SARIS to maximize the average channel power gain of users.
In each transmission, we consider a single-user case, where the target user is randomly distributed in a circular region centered at $\left( {400,0,0} \right)$.
The BS's position is $\left( {0,0,0} \right)$.
The active and passive beamforming in the SARIS system are jointly optimized.
The y-axis of the center of SARIS is set to be 0 to provide the desired average performance for users due to the symmetry of the users' distribution region.
The x-axis and z-axis of the center of SARIS are optimized.

In Fig. 4, the user's average channel power gain is shown versus the x-axis and z-axis of the SARIS's center.
It shows that there exists an optimal 3D position for SARIS.
Intuitively, if SARIS is close to BS or user, the doubled path loss is minimized, but the excessive path loss originated from NLoS connections becomes severe.
Moreover, SARIS's higher altitude makes it easy to establish LoS connections between SARIS and BS/user, but this leads to an increased signal attenuation due to the increase of communication distance.
Therefore, in SARIS deployment, there is a trade-off between doubled path loss and excessive path loss.
Furthermore, there is also a trade-off between path loss and NLoS probability.
This is different from TRIS's deployment, where TRIS's optimal deployment is to place the TRIS close to BS or user.

\subsection{Achievable Rate Enhancement}
Next, we examine the user's achievable rate in the SARIS system in Fig. 5.
Fig. 5 (a) shows that the achievable rate increases with the number of UAVs (i.e., $L$) where $R_A = 10 \text{ m}$, $R_U = 100 \text{ m}$, which demonstrates the increased aperture gain resulting from multiple UAVs.
Furthermore, the achievable rate is significantly improved by optimizing SARIS's deployment. 
The achievable rate improvement is notable when the user is far away from the BS (i.e., $x_U$ is large).
In the case without SARIS's deployment optimization, we suppose that the center of SARIS is at the altitude of 50 m above the center of ground users.

In Fig. 5 (b), we compare the user's achievable rate versus the radius of UAV swarm (i.e., $R_A$) under different radii of ground users (i.e., $R_U$), where SARIS's deployment is optimized, $x_U = 200 \text{ m}$, and $L = 10$.
It is interesting to observe that the achievable rate decreases with $R_A$.
This is because we optimize the UAV swarm center in 3D space, and UAV's apparent deviation from the optimal position will impact the system performance.
Based on the above analysis, we have a design insight of reducing $R_A$ in SARIS.
In addition, users' distribution also impacts system performance.
When $R_U$ increases, the user's average achievable rate decreases due to the increased communication distance.

\section{Open Research Opportunities}
In this section, we briefly illustrate some potential but significant research directions in SARIS-assisted wireless networks.

\subsection{Integrating SARIS in Emerging Applications}
Various wireless applications and scenarios arise in beyond-5G and 6G networks, such as mobile edge computing, wireless power transfer, symbiotic radio system, and physical layer security.
Integrating SARIS in emerging applications with advanced network planning and design insights is essential for achieving smart and intelligent wireless ecosystems.
Assorted resources, including communication, computation, energy, reflecting elements, and trajectory resources, can be optimized to further improve system performance, from different aspects like spectral efficiency, energy efficiency, fairness, secrecy rate, etc.
Efficient communication protocols for cooperative SARIS can be designed for a targeted set of applications.

\subsection{Channel Modeling for ARIS/SARIS}
The practical channel model for ARIS/SARIS remains unknown.
Existing works utilized the 3rd Generation Partnership Project and International Telecommunication Union defined conventional channel models that simplify the array response on RIS as diagonal matrices.
However, these channel models lack the support of measured data in practical engineering.
A more sophisticated channel model needs to consider the impacts of various fading factors, channel correlation, polarization direction, incident angle, operating frequency, etc~\cite{9206044}.
In addition, the impact of UAV wobbling on the cascaded channel is also crucial to be characterized for designing robust beamforming algorithms.

\subsection{Air-Ground Integrated RIS}
As aerial networks enjoy flexible mobility and deployment, they provide reliable wireless connections for ground nodes.
Terrestrial networks contain heterogeneous infrastructures and have been widely developed in recent decades.
In future wireless networks, RIS can be implemented in air-ground integrated networks to improve system performance further.
For example, multi-hop RIS-assisted wireless communications between TIRS and SARIS can be designed with extended coverage and improved capacity.
Moreover, the intelligent surface not only serves as a reflector but also can be equipped at the transmitter or receiver for beamforming, which motivates multiple RIS in a holistic air-ground integrated network.

\subsection{Robust Beamforming Design for SARIS}
Due to UAV wobbling and the difficulty of acquiring accurate UAVs' positions, it would be challenging to realize perfect CSI in practical engineering.
Thus, the cascaded channel estimation error is inevitable.
It is imperative to develop robust beamforming designs for SARIS while guaranteeing users' QoS requirements under UAVs' channel estimation errors.
Moreover, the bounded and statistical CSI error models can be studied under different emerging applications with targeted QoS requirements.

\section{Conclusions}
This article provides an overview of SARIS for achieving reliable and cooperative communication with flexible deployment in future wireless networks.
Notably, SARIS can achieve increased aperture gain by adjusting the reflection coefficients on UAVs.
Based on the LoS and NLoS connections of large-scale path loss, this article demonstrates that the system performance is improved with the advanced 3D deployment design for SARIS.
As SARIS leads to a complete paradigm shift in current terrestrial wireless networks, this article presents some transformative SARIS applications in wireless networks, discusses the challenges of designing SARIS, and illustrates potential research opportunities.
It is hoped that this article would provide a practical guide for integrating RIS in future air-ground wireless networks.

\bibliographystyle{IEEEtran}
\bibliography{IEEEabrv,my}

\section{Biographies}
\small
\noindent
\textbf{Bodong Shang} received his M.S. degree in communication and information system from Xidian University, Xi'an, China, in 2018. He is currently pursuing the Ph.D. degree with the Bradley Department of Electrical and Computer Engineering, Virginia Tech, Blacksburg, USA. His current research interests include several aspects of wireless communications, such as unmanned aerial vehicle, reconfigurable intelligent surface, and edge computing.

\vspace{0.2cm}
\noindent
\textbf{Rubayet Shafin} 
received his Ph.D. degree from Virginia Tech, USA. He is currently working as a senior research engineer with Samsung Research America. His research interests include Beyond-5G/6G cellular networks and machine learning/deep learning for wireless communication.

\vspace{0.2cm}
\noindent
\textbf{Lingjia Liu} received his Ph.D. degree in Electrical and Computer Engineering from Texas A\&M University, USA and his B.S. in Electronic Engineering from Shanghai Jiao Tong University. Currently, he is an Associate Professor in the Bradley Department of Electrical and Computer Engineering and is serving as the Associate Director of Wireless@VT at Virginia Tech, USA. His research interests include machine learning for wireless communications, enabling technologies for 5G and beyond, mobile edge computing, and Internet of Things.

\end{document}